\documentclass[aps,nofootinbib,preprintnumbers,tightenlines]{revtex4}
\usepackage{epsfig}
\begin{document}
\title{Localized tadpoles and anomalies in 6D orbifolds\footnote{Talk given 
at International Conference on Flavor Physics, KIAS, Seoul, 6-11 Oct 2003}}
\author{Hyun Min Lee\footnote{Email: minlee@th.physik.uni-bonn.de}}
\affiliation{Physikalisches Institut, Universit\"at Bonn \\ 53115 Bonn, Germany
}
\date{\today}

\begin{abstract}
In this talk, we review the instability due to radiatively induced FI tadpoles 
in ${\cal N}=2$ supersymmetric gauge theories on orbifolds in six dimensions.
Even with the localized FI tadpoles, we have the unbroken supersymmetry 
at the expense of having the spontaneous localization of bulk zero mode. 
We find the non-decoupling of massive modes unlike the 5D case. 
We also comment on the local anomaly cancellation.
\end{abstract}

\maketitle

\section{Introduction and summary}
Recently models with extra dimensions have drawn a great attention
from particle physicists due to the interesting possibility 
that all or part of the Standard Model particles are
confined to the hypersurfaces in higher dimensions, the so called branes. 
We call the orbifold
fixed points of extra dimensions also the branes. 
When we have in mind the embedding of field theoretic orbifolds
into the string or M-theory, they often contain the remnant supersymmetry
from higher dimensional supersymmetry 
which could describe the supersymmetric standard model 
in four dimensions at low energies.

In the $d=4$ supersymmetric theory with a $U(1)$ factor, it is known that
the Fayet-Iliopoulos term(FI) 
can be radiatively generated for the nonvanishing 
sum of $U(1)$ charges\cite{nilles}.
The FI term in $d=4$ could introduce the quadratic divergence 
even in the supersymmetric theory for not breaking the anomalous $U(1)$.
However, the situation is somewhat different in orbifold models.
In this case, the FI term can be radiatively generated 
at the orbifold fixed points and pose the instability problem 
in a different way. 
With globally vanishing but locally nonzero FI term on orbifolds in $d=5$, 
the supersymmetry condition gives rise to the dynamical localization
of the bulk zero mode and the heavy massive modes\cite{nibbel5d1,nibbel5d2}.
This could open a new possibility for explaining the fermion mass
hierarchy and other scale problems in particle physics with some overlap
of wave functions.
The presence of localized FI terms is consistent  
up to the introduction of 
a bulk Chern-Simons term to cancel the locally nonvanishing 
$U(1)$ mixed gravitational 
anomalies\cite{ah,serone1,barbieri,pilo,kkl,nibbel5d2}. 

In this talk, we examined the more complicated case of co-dimension 2 in the
framework of an ${\cal N}=2$ supersymmetric orbifold theory in $d=6$\cite{lnz}. 
The co-dimension 2 case
should be more relevant for the discussion of compactified superstring theories
in $d=10$ where we have 3 complex extra dimensions. 
We find a localization phenomenon of the bulk zero mode but the situation
differs from the co-dimension 1 case in the sense that the bulk field retains
its six-dimensional nature. The spectrum of massive modes turns out to
be equivalent to a spectrum in the presence of a constant Wilson line.
The potential problem of localized gauge or gravitational anomalies is cured 
with the help of a generalized Green-Schwarz mechanism.

\section{Setup of $d=6$ supersymmetric orbifold}
We consider a $d=6$ ${\cal N}=2$ supersymmetric $U(1)$ gauge theory 
compactified on an orbifold $T^2/Z_2$. Due to the absence of central charge 
in the $d=6$ supersymmetry algebra, there is no off-shell formulation possible
for hypermultiplets.  
The $d=6$ supersymmetry has $SU(2)_R$ as the automorphism group. 

First let us construct the bulk Lagrangian for the vector multiplet and the
hypermultiplets. 
The {\it off-shell} vector multiplet consists of the gauge boson 
$A_M(M=0,1,2,3,5,6)$, the $SU(2)_R$ doublet gaugino $\Omega^i(i=1,2)$
and the $SU(2)_R$ triplet auxiliary field ${\vec D}=(D_1,D_2,D_3)$.  
The gaugino is subject to the right-handed symplectic Majorana-Weyl 
conditions\footnote{The metric convention is $\eta_{MN}={\rm diag}(+-----)$.
The gamma matrices in $d=6$ are $8\times 8$ matrices,
$\Gamma^M=\left(\begin{array}{ll}0&\gamma^M \\ {\bar\gamma}^M&0
\end{array}\right)$ with $\gamma^M=(\gamma^a,\gamma^5,-{\bf 1}_4)$
and ${\bar\gamma}^M=(\gamma^a,\gamma^5,{\bf 1}_4)$ where $a=0,1,2,3$
are four-dimensional indices. The $d=6$ chirality operator
is given by $\Gamma^7=-\tau^3\otimes {\bf 1}_4$ and the charge conjugation
is $C=-i\tau^2\otimes {\cal C}$ where $\cal C$ is the $d=5$ 
charge conjugation.} such as
${\bar\Omega}_i=\varepsilon_{ij}(\Omega^j)^TC$ 
and $\Gamma^7\Omega^i=\Omega^i$. 
The Lagangian for the abelian vector multiplet is given by
\begin{equation}
{\cal L}_V=-\frac{1}{4}F_{MN}F^{MN}
+i\bar{\Omega}\Gamma^M\partial_M\Omega+\frac{1}{2}\vec{D}^2\label{symact}
\end{equation}
which is invariant under supersymmetry transformation:
\begin{eqnarray}
\delta A_M & = &i\bar{\varepsilon}\Gamma_M\Omega\\
\delta \Omega & = & \frac{1}{4}\Gamma^{MN}\varepsilon F_{MN}-\frac{i}{2}\vec{\tau}\varepsilon\vec{D}\\
\delta\vec{D} & = & \bar{\varepsilon}\vec{\tau}\Gamma^M\partial_M\Omega.
\end{eqnarray}
Here the supersymmetry parameter $\varepsilon$ is also a right-handed 
symplectic Majorana-Weyl fermion, satisfying a similar relation as the gaugino.

On the other hand, the {\it on-shell} $r$ hypermultiplets 
contain $SU(2)_R$ doublet scalars $h^\alpha_i(\alpha=1,\cdots, 2r)$ 
and $SU(2)_R$ singlet hyperinos $\zeta^\alpha(\alpha=1,\cdots, 2r)$.
The scalars $h^\alpha_i$ satisfy a reality condition, 
$h^i_\alpha\equiv h_i^{\alpha *}=\varepsilon^{ij}\rho_{\alpha\beta}h_j^\beta$
with $\rho={\bf 1}\otimes\varepsilon$, and the supersymmetry transformation 
laws give constraints to hyperinos: 
$\bar{\zeta}_\alpha = -\rho_{\alpha\beta}{\zeta^{\beta}}^TC$ 
and $\Gamma^7\zeta^{\alpha}=- \zeta^{\alpha}$ which tells that the hyperinos 
have opposite chirality to the one of the gaugino.
Then the Lagrangian for the hypermultiplets is
\begin{eqnarray}
{\cal L}_H={\rm tr}\bigg[\frac{1}{2}|{\cal D}_M h^\alpha|^2
+\frac{i}{2}\bar{\zeta}_\alpha\Gamma^M{\cal D}_M\zeta^\alpha
-2ig\bar{\zeta}_\alpha Q^\alpha{}_\beta \Omega h^\beta
+\frac{i}{2}h^\dagger_\alpha Q^\alpha{}_\beta h^\alpha
({\vec\tau}\cdot{\vec D})\bigg]
\label{hyperact}
\end{eqnarray}
where ${\rm tr}$ is the trace over $SU(2)_R$ indices, 
${\cal D}_M h^\alpha_i=\partial_M h_i^\alpha-g A_M Q^\alpha{}_\beta h_i^\beta$
and $Q=-iq\otimes \tau_3$ with $q$ being $U(1)$ charge matrix.
This Lagrangian is invariant under supersymmetry transformation up to the 
equations of motion:
\begin{eqnarray}
\delta h_i^\alpha & = &i\bar{\varepsilon}_i\zeta^\alpha\\
\delta \zeta^\alpha & = & -\Gamma^{A}\varepsilon^i {\cal D}_A h_i^\alpha.
\end{eqnarray}

We now consider our theory given by the sum of the Lagrangians (\ref{symact})
and (\ref{hyperact}) on the orbifold $T^2/Z_2$.
The coordinates for the torus are $x^5$ and $x^6$ with radii $R_5$
and $R_6$, respectively.
The $Z_2$ action on coordinate space is defined as 
\begin{eqnarray}
Z_2:~~ (x^5,x^6)\to (-x^5,-x^6).
\end{eqnarray}
So there are four fixed points on $T^2/Z_2$ orbifold:
$(x^5,x^6)=(0,0),(\pi R_5, 0),(0,\pi R_6),\,\,{\rm and}\,\, (\pi R_5,\pi R_6)$.
Then the orbifold action on the field space can be read from the Lagrangian 
as the following: for the vector multiplet, 
\begin{eqnarray}
A_m(-x^5,-x^6)&=&A_m(x^5,x^6), \ \ \ A_{5,6}(-x^5,-x^6)=-A_{5,6}(x^5,x^6), 
\label{bc1}\\
{\vec\tau}\cdot{\vec D}(-x^5,-x^6)
&=&\tau_3({\vec\tau}\cdot{\vec D}(x^5,x^6))\tau_3, \label{bc2}\\
\Omega(-x^5,-x^6)&=&-i({\bf 1}\otimes \tau_3)\Gamma_5\Gamma_6\Omega(x^5,x^6),
\label{bc3} 
\end{eqnarray}
and for the hypermultiplets,
\begin{eqnarray}
h(-x^5,-x^6)&=&-\eta({\bf 1}\otimes \tau_3)h(x^5,x^6)\tau_3, \label{bc4}\\
\zeta(-x^5,-x^6)&=&i\eta({\bf 1}\otimes \tau_3)\Gamma_5\Gamma_6\zeta(x^5,x^6)
\label{bc5}
\end{eqnarray}
where $\eta=\pm 1$.
Defining a $d=6$ spinor such as $\psi=(\psi_L,\psi_R)$
in the four-dimensional Weyl representation, 
the orbifold action for the gaugino and the hyperinos  
becomes respectively
\begin{eqnarray}
\Omega_R(-x^5,-x^6)&=&i({\bf 1}\otimes \tau_3)\gamma^5\Omega_R(x^5,x^6), 
\label{bc3a}\\
\zeta_L(-x^5,-x^6)&=&i\eta({\bf 1}\otimes \tau_3)\gamma^5\zeta_L(x^5,x^6)
\label{bc5a}
\end{eqnarray} 
where the subscripts $R,L$ denote the $d=6$ chiralities.

We can solve the reality conditions by 
\begin{eqnarray}
\Omega^i_R=\left(\begin{array}{l}\,\,\chi 
\\ {\cal C}{\bar\chi}^T\end{array}\right), \ \
h^\alpha_i=\left(\begin{array}{ll}\,\,\,\,\phi^{*{\hat\alpha}}_- & 
\phi^{\hat\alpha}_+
\\ -\phi^{*{\hat\alpha}}_+ & \phi^{\hat\alpha}_- \end{array}\right), \ \
\zeta^\alpha_L=\left(\begin{array}{l}\,\,\psi^{\hat\alpha} \\ 
{\cal C}(\bar{\psi^{\hat\alpha}})^T\end{array}\right)
\end{eqnarray}
where $\hat\alpha=1,\cdots, r$. 
Then, after applying the orbifold boundary conditions with $\eta=+1$ 
on those redefined fields,
the parities of all bulk fields are collected in the Table 1
where use is made of $d=4$ chiral projection on the Majorana spinors $\chi_\pm$ 
and $\psi^{\hat\alpha}_{\pm}$ for the gaugino and the hyperinos, respectively.
For instance, the four-component gaugino $\chi$ is written in terms of 
two-component Weyl spinors of $\chi_\pm$ as $\chi=\chi_{+L}-\chi_{-R}$ and 
${\cal C}{\bar\chi}^T=\chi_{+R}+\chi_{-L}$. 
\begin{table}
\begin{center}
V:~~ \begin{tabular}{|c|c|c|c|c|c|c|}
\hline
Field & $A_m$ & $A_{5,6}$ & $D_{1,2}$ & $D_3$ & $\chi_{\pm L}$ & 
$\chi_{\pm R}$\\
\hline
Parity & $+1$ & $-1$ & $-1$ & $+1$ & $\pm 1$ & $\pm 1$ \\
\hline
\end{tabular}
\qquad
H:~~ \begin{tabular}{|c|c|c|c|}
\hline
Field & $\phi^{\hat\alpha}_\pm$ & $\psi^{\hat\alpha}_{\pm L}$ 
& $\psi^{\hat\alpha}_{\pm R}$ \\
\hline
Parity & $\pm 1$ & $\pm 1$ & $\pm 1$ \\
\hline
\end{tabular}
\end{center}
\caption{Parities of vector and hyper multiplets}
\end{table}
As a result, a four-dimensional ${\cal N}=1$ vector multiplet at the fixed 
points is composed of
\begin{eqnarray}
(A_m,\chi_{+L},-D_3+F_{56}), \label{branev}
\end{eqnarray}
i.e. the four-dimensional auxiliary field is not $D_3$
as one might have naively expected but $-D_3+F_{56}$ 
which is the gauge covariant generalization of $-D_3+\partial_5 \Phi$ 
in $d=5$\cite{peskin,nibbel5d1,nibbel5d2}.
Therefore, we can couple chiral multiplets which live
at the fixed points to the ${\cal N}=1$ vector multiplet (\ref{branev}).
Then, the total Lagrangian contains
\begin{eqnarray}
{\cal L}_{bulk}&=&\sum_{\pm}(|{\cal D}_M\phi_\pm|^2
\mp g\phi^\dagger_\pm q\phi_\pm D_3)+i{\bar\psi}{\bar\gamma}^M{\cal D}_M\psi
+\cdots, \label{rel1}\\
{\cal L}_{brane}&=&\sum^4_{I=1}\delta(x^5-x^5_I)\delta(x^6-x^6_I)
[|D_m\phi_I|^2+g\phi^\dagger_I q_I \phi_I(-D_3+F_{56})+\cdots]\label{rel2}
\end{eqnarray}
where we omitted $\hat\alpha$ indices in the bulk Lagrangian, 
$(x^5_I, x^6_I)$ label the fixed points and 
$q_I$ are the charge matrices at the fixed points.

\section{Fayet-Iliopoulos tadpoles}
As discussed in the previous section, 
the $D$ field belonging to the four-dimensional vector 
multiplet is given by $D=-D_3+F_{56}$.
So the form of our Fayet-Iliopoulos(FI) term is
\begin{eqnarray}
{\cal L}_{FI}=\xi(-D_3+F_{56}).
\end{eqnarray}
The coefficient of FI term can be computed by considering eqs.~(\ref{rel1})
and (\ref{rel2}) with the standard procedure 
as in $d=5$\cite{nibbel5d1,nibbel5d2}. 
The sum of bulk and brane contributions to the FI term is
\begin{eqnarray}
\xi=\sum_I\left(\xi_I+\xi''(\partial_5^2+\partial_6^2) \right)\delta(x^5-x^5_I)\delta(x^6-x^6_I)
\end{eqnarray}
with
\begin{eqnarray}
\xi_I&=&\frac{1}{16\pi^2}g\Lambda^2\left(\frac{1}{4} {\rm tr}(q)+ {\rm tr} (q_I)\right),\\
\xi''&=&\frac{1}{16}\frac{1}{16\pi^2}g\ln \Lambda^2 {\rm tr} (q).
\end{eqnarray}
We note that the bulk contribution has both quadratically divergent 
and logarithmically divergent pieces which are equally distributed 
at the fixed points 
whereas the brane contribution has only the quadratic divergence.  

Then, from the effective potential with the FI tadpoles\cite{lnz}, 
we find the conditions for unbroken supersymmetry
\begin{equation}
\langle D_3\rangle=\langle F_{56}\rangle
=g(\langle\phi_+\rangle^{\dagger}q\langle\phi_+\rangle
-\langle\phi_-\rangle^{\dagger} q\langle\phi_-\rangle)+\xi
+g\sum_I\delta(x^5-x^5_I)\delta(x^6-x^6_I)
\langle\phi_I\rangle^\dagger q_I\langle\phi_I\rangle\label{cond1}
\end{equation}
together with
\begin{eqnarray}
\langle\phi_+\rangle^Tq\langle\phi_-\rangle=0 \ \ \ {\rm and}
\ \ \ \langle({\cal D}_5+i{\cal D}_6)\phi_\pm\rangle=0.
\end{eqnarray}
Provided that the ground state does not break the $U(1)$, 
i.e. $\langle \phi_\pm\rangle=\langle \phi_I\rangle=0$, 
the supersymmetry condition (\ref{cond1}) becomes
\begin{equation}
\langle F_{56}\rangle=\xi.\label{back}
\end{equation}
After integrating both sides over the extra dimensions, the Stokes theorem 
with no boundary tells
\begin{eqnarray}
\sum_I \xi_I=0 \ \ \ {\rm or} \ \ \ 
{\rm tr}(q)+(q_1)+(q_2)+(q_3)+(q_4)=0\label{sum}. 
\end{eqnarray}
This consistency condition ensures the absence
of overall mixed gauge-gravitational anomalies. 
If (\ref{sum}) is
violated, we would expect the U(1) to be broken at a high scale
either spontaneously or through a variant of Green-Schwarz 
mechanism\cite{green}.
Even if the consistency condition is satisified, 
we need the local version of Green-Schwarz mechanism for the local
anomaly cancellation\cite{serone,antoniadis,walter,quiros,lnz}. As a result, 
the $U(1)$ is broken or unbroken depending on whether the dual axion 
lives on the brane or in the bulk\cite{serone,antoniadis,walter,quiros,lnz}.

Then, taking the following 
ansatz\footnote{Here we have fixed the gauge implicitly}
\begin{equation}
\langle A_5\rangle=-\partial_6 W\qquad
\mbox{and}\qquad \langle A_6\rangle=\partial_5 W,\label{W}
\end{equation}
eq.~(\ref{back}) becomes a sort of Poisson 
equation\footnote{We are using complex coordinates 
$z=\frac{1}{R_5}x^5+\frac{1}{R_6}\tau x^6$ 
with the torus modulus $\tau=iR_6/R_5$. The periodicities on the torus then 
are $z\simeq z+2\pi\simeq z+2\pi\tau$.}
\begin{equation}
\partial\bar{\partial}W'=\sum_I\xi_I\delta^2(z-z_I),\label{Poisson}
\end{equation}
with
\begin{equation}
W'=2\left(W-\frac{2}{R^2_5}\sum_I\xi''\delta^2(z-z_I)\right).\label{def2}
\end{equation}
Consequently, the solution to eq.~(\ref{Poisson}) is given by the propagator
of a bosonic string for a toroidal world sheet as
\begin{equation}
W'=\frac{1}{2\pi}\sum_I\xi_I
\left[\ln\left|\vartheta_1\left(\frac{z-z_I}{2\pi}\big|
\tau\right)\right|^2-\frac{1}{2\pi\tau_2}[{\rm Im}(z-z_I)]^2\right]
\label{sol}
\end{equation}
where $\tau_2={\rm Im}\tau=R_6/R_5$. 
Note that in order for $W'$ in the above to be a solution
to (\ref{Poisson}), eq. (\ref{sum}) must hold.

\section{Localization of the bulk zero mode and mass spectrum}
As shown before, in the presence of localized FI tadpoles, 
the unbroken gauge symmetry and supersymmetry requires
the nontrivial profile of the extra dimensional components 
of gauge field. This nontrivial background modifies the wave functions 
and the mass spectrum of bulk modes. 

First let us consider the solution for the zero mode in the presence of the
background solution. The equation for the zero mode is
\begin{eqnarray}
({\bar \partial}-gq{\bar \partial}W)\phi_+=0
\end{eqnarray}
where $A=A_5-iA_6=-(2i/R_5)\partial W$ in the complex coordinates.
Thus, we find the exact solution for the zero mode as
\begin{eqnarray}
\phi_+&=&f_+ e^{gqW} \nonumber \\
&=&f_+\prod_I \left|\vartheta_1\left(\frac{z-z_I}{2\pi}\big|
\tau\right)\right|^{\frac{1}{2\pi}gq\xi_I}\times \nonumber \\
&\times&\exp\left[-\frac{1}{8\pi^2\tau_2}gq\xi_I[{\rm Im}(z-z_I)]^2
+ \frac{gq\xi''}{R^2_5}\delta^2(z-z_I)\right]
\label{final}
\end{eqnarray}
where $f_+$ is a complex integration constant which is determined
by the normalization condition
\begin{eqnarray}
1=\int_0^{\pi R_5}dx^5\int_0^{\pi R_6}dx^6|\phi_+|^2.
\end{eqnarray}
Therefore, from the asymptotic limit of the theta function 
\begin{eqnarray}
\vartheta_1\bigg(\frac{z-z'}{2\pi}|\tau\bigg)\rightarrow
(\eta(\tau))^3 (z-z') \ \ \ {\rm for} \ \ z\rightarrow z'
\end{eqnarray}
where $\eta(\tau)$ is the Dedekind eta function, 
the $\vartheta_1$ term shows the similar tendency for localization 
as the $e^{({\rm Im})^2}$ term but it would mean a strong localization 
of the zero mode due to the divergence at the fixed point(s) with $q\xi_I<0$.
Moreover, the $e^{\delta^2}$ term also seems to give a strong (de)localization
for $q\xi''>0(q\xi''<0)$ as in the five-dimensional case \cite{nibbel5d2}.
To understand the localization of the zero mode explicitely we have to
maintain two regularization scales: the momentum cutoff $\Lambda$ and the
brane thickness $\rho$; both $\rho$ and $1/\Lambda$ are small compared to
$R_5, R_6$. The localization induced by $\xi$ is typically exponential
in $\Lambda$ while the one induced by $\xi''$ is power like. Thus as
long as $\rho$ is not very small compared to $1/\Lambda$ the effect of the
logarithmic FI-term will be subleading (naturally one could expect
$\rho$ and  $1/\Lambda$ to be of the same order of magnitude).

Next let us consider the equation for the massive modes
with nonzero gauge field background 
\begin{equation}
(\partial\pm gq\partial W)(\bar{\partial}\mp gq\bar\partial W)\phi_\pm=-\frac{1}{4}m^2R^2_5\phi_\pm.\label{eom1}
\end{equation}
By substituting in eq. (\ref{eom1})
\begin{eqnarray}
\phi_\pm = e^{\pm gqW}\tilde{\phi}_\pm,
\end{eqnarray}
we get a simpler form
\begin{eqnarray}
\partial\bar{\partial}\tilde{\phi}_\pm\pm 2gq\partial W\bar{\partial}
\tilde{\phi}_\pm=-\frac{m^2}{4}R_5^2\tilde{\phi}_\pm.\label{eom2}
\end{eqnarray}
Since there appear derivatives of delta functions in this equation,
we need to regularize the delta function.
Let us take the regularizing 
function\footnote{See, for instance, Ref.~\cite{lnz}.} as 
$\Delta^2(z-z_I)$ which satisfies 
$\lim_{\rho_I\rightarrow 0}\int d^2 z'\Delta^2(z-z')h(z',{\bar z}')=h(z,{\bar z})$ for an arbitrary complex function $h$ 
and is zero only for $|z-z_I|>\rho/R_5$ with $\rho$ being the brane 
thickness. Then, we get the holomorphic derivative of $W$ as
\begin{eqnarray}
\partial W=\frac{1}{2}\sum_I\xi_I\int d^2 z'\,\partial G(z-z')\Delta^2(z'-z_I)
+\frac{2}{R^2_5}\sum_I\xi^{\prime\prime}\partial\Delta^2(z-z_I)\label{pW}
\end{eqnarray}
where $G(z-z')$ is the string propagator on the torus satisfying
${\bar \partial}\partial G(z-z')=\delta^2(z-z')-1/(8\pi^2\tau_2)$.
Then, since $\partial W$ is holomorphic for $|z-z_I|>\rho_I/R_5$,
eq.~(\ref{eom2}) becomes solvable
with a separation of variables
as ${\tilde\phi}_{\pm}=\chi_\pm(z)\,\varphi_{\pm}({\bar z})$.
After all, imposing the periodicity on $\tilde\phi_\pm$\cite{lnz}, 
we find the mass spectrum 
\begin{eqnarray}
m^2=\frac{4}{R^2_5}\bigg|c_\pm
\pm\frac{gq}{8\pi^2\tau_2}\sum_I\xi_I{\bar z}_I\bigg|^2, \label{mass}
\end{eqnarray}
with
\begin{eqnarray}
c_\pm=\frac{1}{2}\bigg(\frac{n'}{\tau_2}+in\bigg),
\end{eqnarray}
where $n$ and $n'$ are integers.
The structure of the resulting mass spectrum is so different 
from the $d=5$ case 
in the sense that there also appear linear terms in $\xi_I$'s.
In particular, for $\sum_I\xi_Iz_I=0$,
which is the case with no net dipole moments
coming from FI terms, even the nonzero localized FI terms do not modify
the mass spectrum at all.
Even for $\sum_I\xi_Iz_I\neq 0$ with large FI terms,
there generically appears a normal KK tower of massive modes 
starting with large integers $n$ and $n'$ which cancel the shift 
due to local FI terms.

Comparing this mass spectrum with the one obtained in the
five-dimensional case, we observe a qualitative difference.
There, the Kaluza-Klein excitations of the bulk mode became very
heavy with the cut-off $\Lambda$ and in the limit
$\Lambda\rightarrow\infty$ we just retained a massless zero mode
localized at a fixed point. Effectively the bulk field underwent
a dimensional transmutation and became a brane field. In the present
six-dimensional case such a radical effect does not happen. The zero
mode bulk field shows a localization behaviour as illustrated 
in Ref.~\cite{lnz} 
but the Kaluza-Klein excitations are not removed and the
bulk field retains its six-dimensional nature.

\begin{acknowledgments}
The anthor thanks Prof. E. J. Chun and organizers of ICFP03 for making a good 
ambience and a kind support during the workshop. 
This work is supported by the
European Community's Human Potential Programme under contracts
HPRN-CT-2000-00131 Quantum Spacetime, HPRN-CT-2000-00148 Physics Across the
Present Energy Frontier and HPRN-CT-2000-00152 Supersymmetry and the Early
Universe. The author was supported by priority grant 1096 of the Deutsche
Forschungsgemeinschaft.
\end{acknowledgments}


\end{document}